\documentclass[a4paper,10pt,twoside]{cpc-hepnp}

\usepackage{multicol}
\usepackage{graphicx}
\usepackage{booktabs}
\usepackage{amssymb,bm,mathrsfs,bbm,amscd}
\usepackage[tbtags]{amsmath}
\usepackage{lastpage}

\begin{document}

\fancyhead[c]{\small Submitted to 'Chinese Physics C'} \fancyfoot[C]{\small 010201-\thepage}

\footnotetext[1]{ Supported by The Strategic Priority Research Program of the Chinese Academy of Sciences Grant No. XDA10010500}
\footnotetext[1]{ Supported by National Natural Science Foundation of China (11390384)}

\title{The laboratory measurement of radioactivity purification for $^{212}$Pb in liquid scintillator}

\author{%
      Wei Hu (ºúά)$^{1,2;1)}$\email{huwei@mail.ihep.ac.cn}%
\quad Jian Fang (·½½¨)$^{1;2)}$\email{fangj@mail.ihep.ac.cn}%
\quad Bo-Xiang Yu (Óá²®Ïé)$^{1}$
\quad Xuan Zhang (Õş@)$^{1,2}$
\quad Li Zhou (ÖÜÀò)$^{1}$
\\Xiao Cai (²ÌÐ¥)$^{1}$
\quad Li-Jun Sun (ËïÀö¾ý)$^{1}$
\quad Wan-Jin Liu (ÁõÍò½ð)$^{1}$
\quad Lan Wang (Íõá°)$^{1}$
\quad Jun-Guang Lu (ÂÀ¾ü¹â)$^{1}$
}
\maketitle

\address{%
$^1$ State Key Laboratory of Particle Detection and Electronics, (Institute of High Energy Physics, CAS)Beijing 100049, China\\
$^2$ Institute of High Energy Physics, University of Chinese Academy of Sciences, Beijing 100049, China\\
}

\begin{abstract}
The liquid scintillator (LS) has been widely utilized in the past, running and future neutrino experiments, and requirement to the LS radio-purity is higher and higher. The water extraction is a powerful method to remove soluble radioactive nuclei, and a mini-extraction station had been constructed. To evaluate the extraction efficiency and optimize the operation parameters, a setup to load radioactivity to LS and a laboratory scale setup to measure radioactivity which used $^{212}$Bi-$^{212}$Po-$^{208}$Pb cascade decay were developed. Experiences from laboratory study will be useful to the design of large scale water extraction plants and the optimization of working conditions in the future.
\end{abstract}



\footnotetext[0]{\hspace*{-3mm}\raisebox{0.3ex}{$\scriptstyle\copyright$}2013
Chinese Physical Society and the Institute of High Energy Physics
of the Chinese Academy of Sciences and the Institute
of Modern Physics of the Chinese Academy of Sciences and IOP Publishing Ltd}%

\begin{multicols}{2}

\section{Introduction}

The liquid scintillator plays a very important role in intensity frontier neutrino experiments. The Jiangmen Underground Neutrino Observatory (JUNO) is a multi-purpose neutrino experiment, with the primary scientific goal to determine neutrino mass hierarchy. The neutrino detector is filled with LS of 20 ktons fiducial mass. To suppress the accidental background, as well as to achieve the potential goal of solar neutrino studies, the basic radioactivity contamination requirements to JUNO LS is: 10$^{-15}$ g/g (in this paper, g/g means gram of $^{232}$Th or $^{238}$U per gram of LS) for both $^{238}$U and $^{232}$Th.

The general methods to remove radioactivity contaminations are water extraction, nitrogen stripping and distillation, which are sensitive to soluble nuclei, Rn and insoluble nuclei, respectively. Before the mass production of purified LS, each method should have a prototype and optimized operation parameters. Such as Borexino experiment (a solar neutrino experiment, solvent of LS is trimethylbenzene), which holds the world-record radioactivity contaminations with $^{232}$Th/$^{238}$U of 10$^{-18}$g/g \cite{lab1}, the parameters of large scale purification plants and the prototype are consistent \cite{lab2} \cite{lab3}.

Though these purification methods are efficient for LS of Borexino, they need to be carefully studied for LAB-based LS of JUNO. A water extraction prototype had been constructed at IHEP, Beijing, to validate the prototype and optimize operation parameters. LS radioactivity, before and after purification should be measured. However, the typical U and Th contaminations in LS are 10$^{-13}$ to 10$^{-14}$ g/g, and it is impossible to measure such low radioactivity in the laboratory. A general method is to load radioactive nuclei, such as $^{222}$Rn or $^{220}$Rn, to LS, and purify the LS with the prototype, and measure the purified and un-purified LS with a clean detector.

In this paper, the Rn loading technology, the radioactivity measurement setup, and the water extraction efficiencies, as well as the optimized operation parameters are reported. The efficiency has reached the world average level, indicating the prototype is successfully working, and the optimized parameters are useful to future middle-scale and mass production plants.

\section{Radioactivity measurement setups}
\subsection{$^{220}$Rn loading}

The limit of radioactivity measurement in the laboratory is 10$^{-9}$ g/g. But the natural contamination of LS is 10$^{-13}$ to 10$^{-14}$ g/g, hence it is impossible to measure such low radioactivity in the laboratory. In order to study the effect of purification for LS in laboratory experiments, the only solution is an artificial pollution of the samples with radioactivity. Since powder radio-source does not dissolve in LS and the solubility of liquid radio-source in LS is not high, the general method is to load radon to LS. Because radon is non-polar gas, the solubility of radon in LS is high with 13 times of the radon concentration in air under room temperature \cite{lab4}. Therefore, it is effective to load radioactivity in LS by bubbling radon into LS.

The common used radon is $^{222}$Rn, which is from the $^{238}$U decay chain and has a 3.8 days half-life time. For example, in the Borexino experiment, $^{222}$Rn was loaded to LS and the contamination of its daughter $^{210}$Po was measured. The disadvantage is that it requires months for $^{210}$Po to accumulate to a measurable amount, since its mother $^{210}$Pb¡¯s half-life time is 22.3 years \cite{lab5}. Besides, the long half-life time of $^{210}$Pb would pollute the experiment setups.

Compared with $^{222}$Rn, a better candidate is $^{220}$Rn, which is from the $^{232}$Th decay chain as shown in Fig.1, and has 55 s half-life time. After loading, $^{220}$Rn quickly decays to $^{212}$Pb with 10.6 hours half-life time. The decay of $^{212}$Pb¡¯s daughters $^{212}$Bi and $^{212}$Po are famous cascade decays ($\beta$-$\alpha$ cascade decay), since the half-life time of $^{212}$Po is only about 300 ns. The cascade decay supplies a pair of time correlated signals in our experiments, with high efficiency and extra low background. With $^{220}$Rn loading, the water extraction efficiency is estimated with the nucleus $^{212}$Pb. The 10.6 hours half-life time of $^{212}$Pb is long enough to do extraction and measurement, and it will not cause any contamination to experiment setups. Meanwhile, a large amount of $^{220}$Rn in LS will decay to $^{212}$Pb in a very short period of time, leading to high radioactivity loading efficiency.

\begin{center}
\includegraphics[width=5cm]{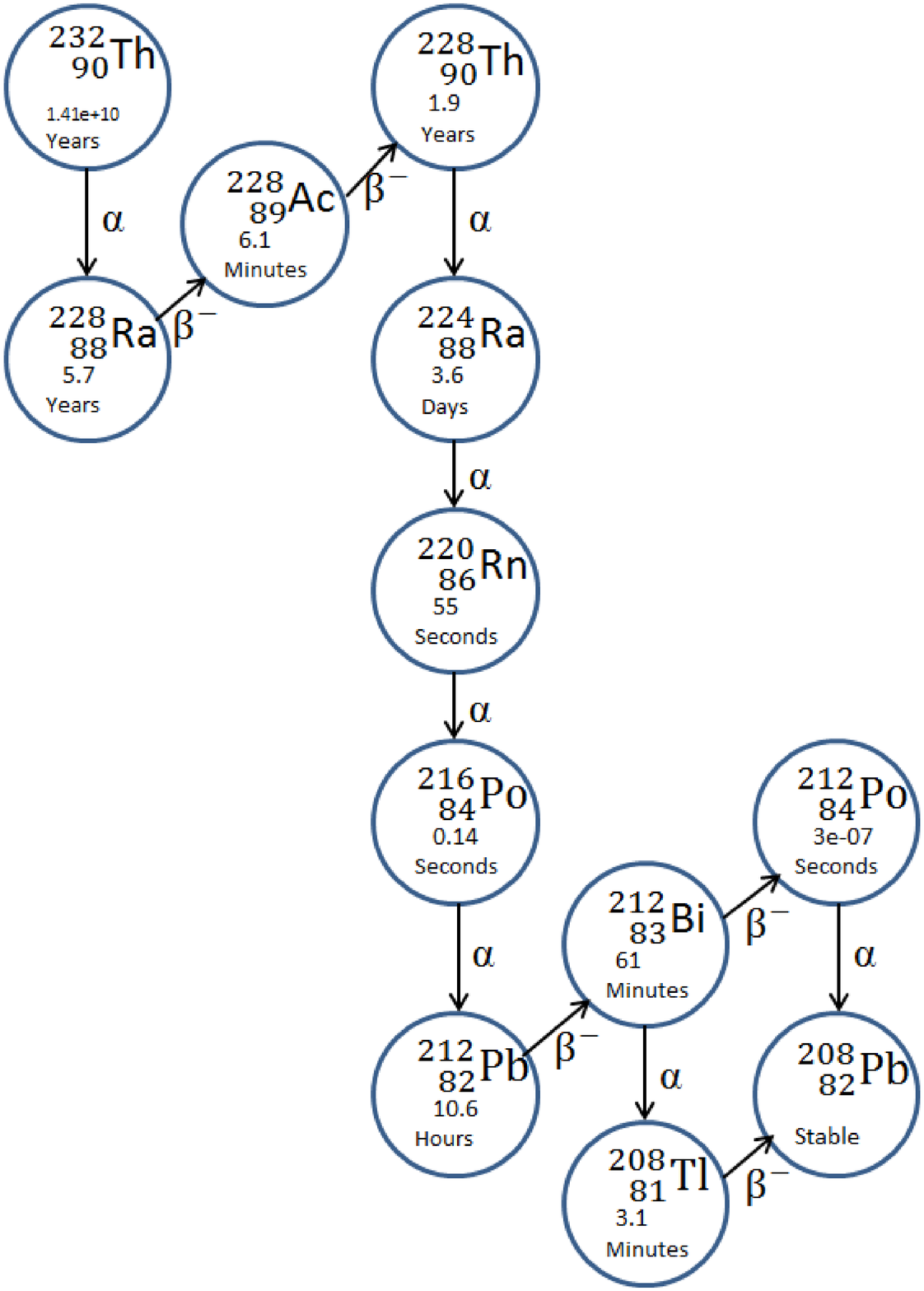}
\figcaption{\label{fig1} The nature decay chain of $^{220}$Rn}
\end{center}

The loading method used in the paper was bubbling $^{220}$Rn into LS sample. The corresponding setup inside a glove box is shown in Fig.~\ref{fig2}: A $^{232}$Th source releasing 1200 Bq $^{220}$Rn produced by Nanhua University was used. Nitrogen went through a bubbler filled with water, then it blew through the source and took $^{220}$Rn out. Finally it went through a bubbler filled with LS. According to the research conducted by Nanhua University, the $^{220}$Rn release rate of $^{232}$Th source increased with higher environment humidity. After bubbling $^{220}$Rn into LS for 74.2 hours, the concentration of $^{212}$Pb reached a balance level. In the following study, LS was bubbled with $^{220}$Rn for 20 hours which reached the 2/3 of the $^{212}$Pb concentration balance level \cite{lab6}.

\begin{center}
\includegraphics[width=7cm]{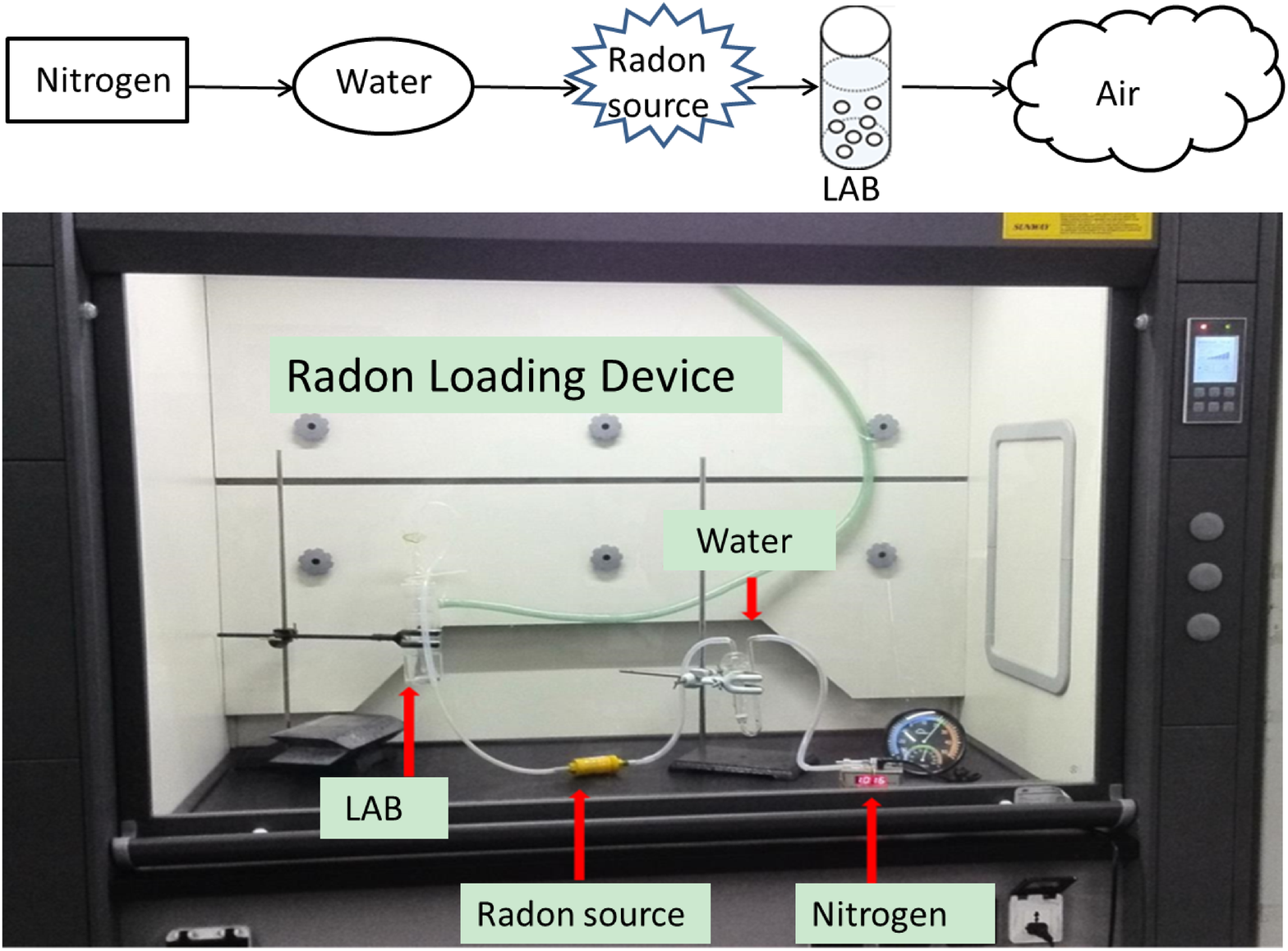}
\figcaption{\label{fig2}   Laboratory setup for radon-loading of LS samples }
\end{center}

The radioactivity was measured by the experimental setup depicted in Fig.~\ref{fig3}. In a light-tight box, a pair of 2¡± PMTs (XP2020) was placed on both sides of an LS sample cell, to do double coincidence measurement. The coincidence measurement could reduce the influence of single PMT's fluctuation to experiment data.  The LS container was a cylinder quartz glass bottle, with 5 cm diameter and 1.5 cm thickness, and a capacity of 17.1 g LS. Gamma rays from ambient radioactivity were attenuated by the shielding of low-activity lead bricks \cite{lab4}. A flash ADC (DT5751 made by CAEN with 1 GHz sampling frequency) was used for data acquisition. The total background event rate (including $\beta$, $\alpha$, $\gamma$ and cascade decay events) during measurement was 0.25 Hz. After finishing the data acquisition for all events, $\beta$-$\alpha$ cascade events were pick out by offline analysis. This setup was designed as a $\beta$-$\alpha$ counting system.

\begin{center}
\includegraphics[width=7cm]{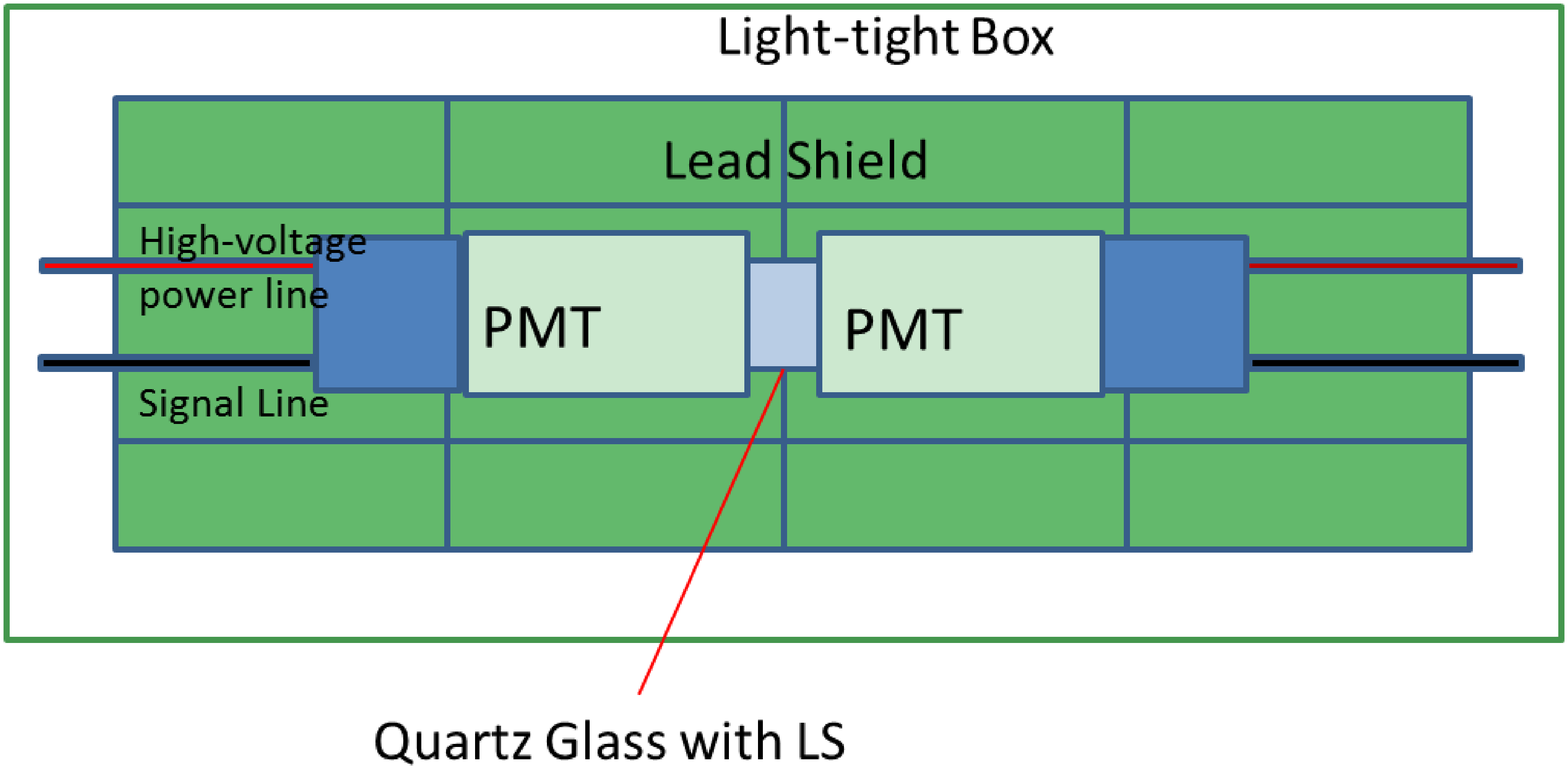}
\figcaption{\label{fig3}   Laboratory setup for measuring the efficiency of radiopurification }
\end{center}

\section{Data analysis}
\subsection{The $\beta$-$\alpha$ cascade event selection}
\subsubsection{Real $\beta$-$\alpha$ cascade event}

After $^{220}$Rn loading, three hours' data was taken to determine the initial $^{212}$Pb concentration. The coincidence time window for signals from the two PMTs was required to be smaller than 5 ns, due to the length difference of the cables connected to the two PMTs. 99.99\% of the $\beta$-$\alpha$ cascade events met this requirement, with the statistical error of 9.55$\times$10$^{-8}$ (statistical error will not be discussed in this section since it was too small).

The time interval distribution between the $\beta$ decay and $\alpha$ decay of the $\beta$-$\alpha$ cascade events is showed in Fig.~\ref{fig4}. The distribution can be described by the formula below~\cite{xuzizong},
\begin{eqnarray}
\label{eq1}
f(t) = \frac{1}{\tau}\times N_0 \times e^{-\frac{t}{\tau}},
\end{eqnarray}

where $\tau$ is the life time of $^{212}$Po and $N_0$ is a parameter related to the concentration of $^{212}$Po. $T_{1/2}$ is the half-life time of $^{212}$Po and $T_{1/2}= ln2*\tau$. Hence, the formula above can be written as,

\begin{eqnarray}
\label{eq2}
f(t) = \frac{ln2}{T_{1/2}}\times N_0 \times 2^{-\frac{t}{T_{1/2}}}.
\end{eqnarray}

The following function was used to fit the time interval distribution.
\begin{eqnarray}
\label{eq3}
f(t) = \frac{1}{p_1} \times p_0 \times 2^{-\frac{t}{p_1}}.
\end{eqnarray}
The parameter $p_1$ in the fitting result stands for the half-life time of $^{212}$Po. In theory, the half-life time is 298 ns, while the experiment result is 298.4$\pm$1.2 ns. Therefore, the $\beta$-$\alpha$ counting setup was reliable for detecting $\beta$-$\alpha$ cascade events.

\begin{center}
\includegraphics[width=7cm]{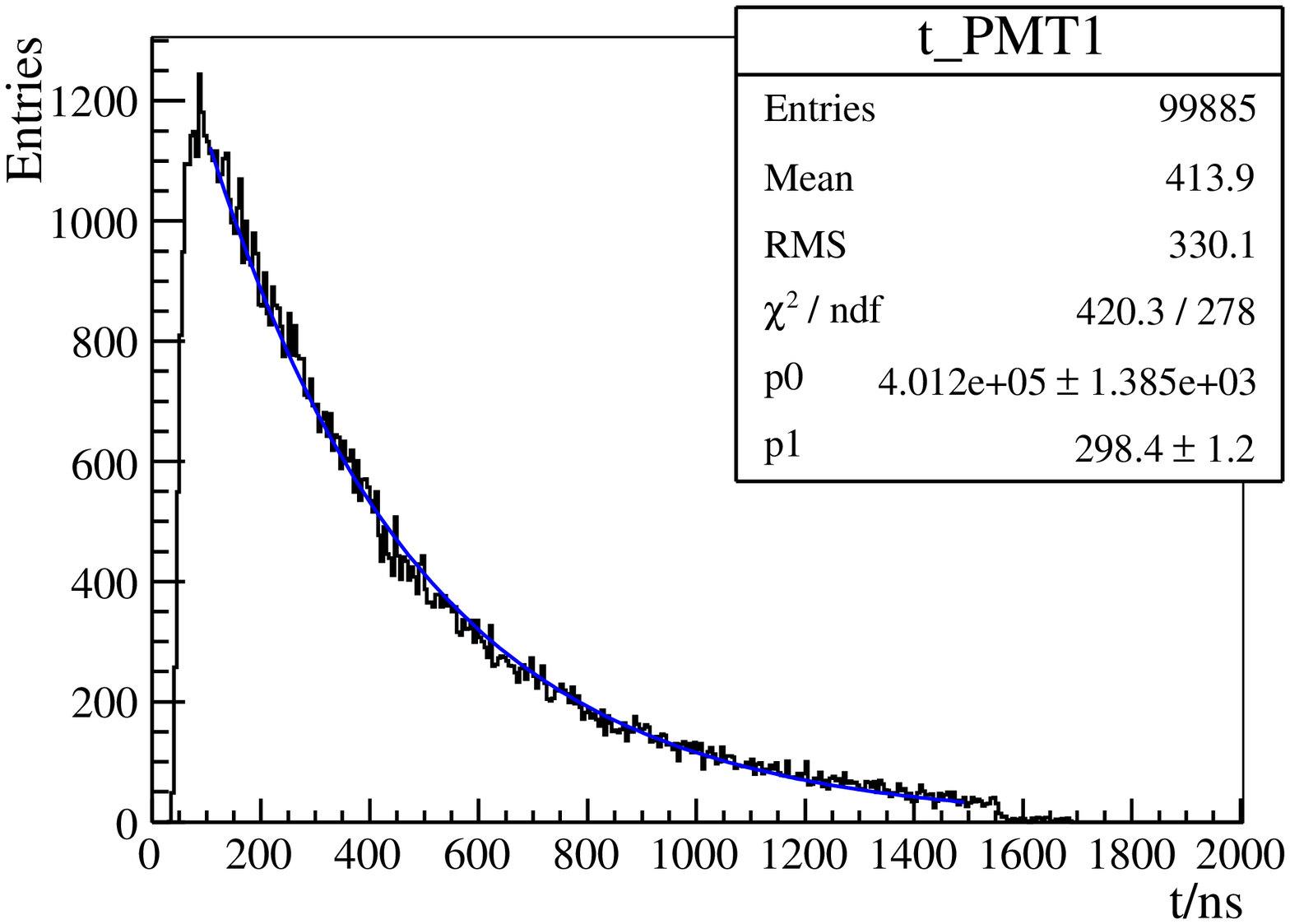}
\figcaption{\label{fig4}   The time interval distribution between the $\beta$ decay and $\alpha$ decay }
\end{center}

For each $\beta$-$\alpha$ cascade event, there was a time interval between the $\beta$ event and the $\alpha$ event. And the time interval measured by the two PMTs should be the same theoretically. At least, the difference must be very small due to the difference of the two PMTs or other impact in experiment. Fig.~\ref{fig5}(a) shows the difference of time interval of $\beta$-$\alpha$ cascade events detected by the two PMTs, which was within the coincidence time window. The entries number before normalization was 104640. Gaussian function was used to fit the distribution and the result was mean value of 0.51 and $\sigma$ of 0.91. The difference of the time interval was almost within (-2 ns, 3 ns). 95.92\% of the events met this requirement, which was consistent to the probability of variable from Gaussian distribution locating within 2 sigma range.

\begin{center}
\includegraphics[width=7cm]{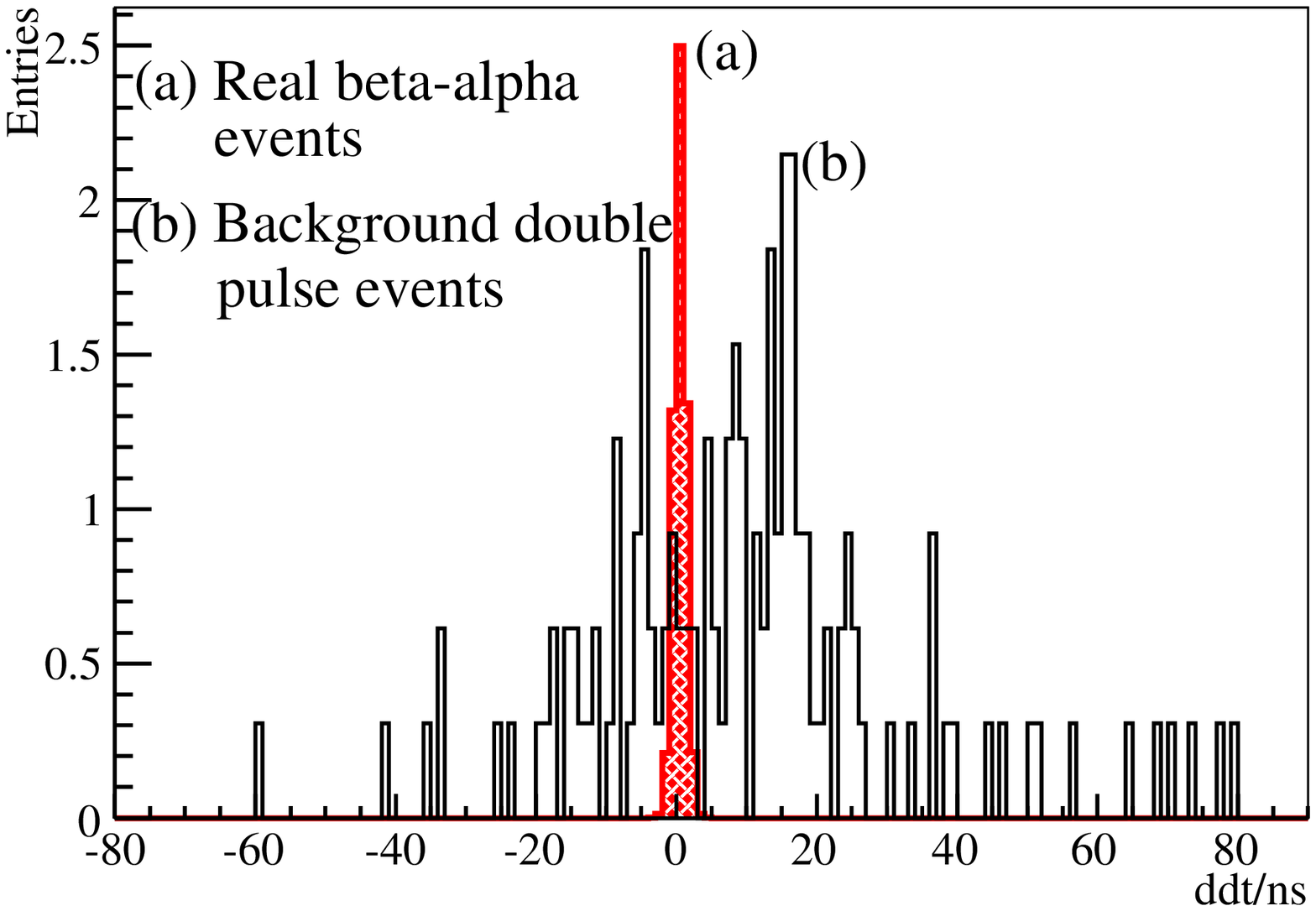}
\figcaption{\label{fig5}   The difference of time interval of $\beta$-$\alpha$ pulse between the two PMTs (The entries are normalized)}
\end{center}

The event energy was proportional to the total charge collected. Measurement of the total charge from PMT was estimated by integrating the entire pulse. Fig.~\ref{fig6}(a) shows the integral value of the $\alpha$ event pulse (the second pulse in a double-pulse event).  Due to the mono-energy of $\alpha$ event, the distribution was centralized like a Gaussian distribution. 92.80\% of the second pulse integral value was within (1700 FADC, 8000 FADC) after the two selection criterions discussed above.

\begin{center}
\includegraphics[width=7cm]{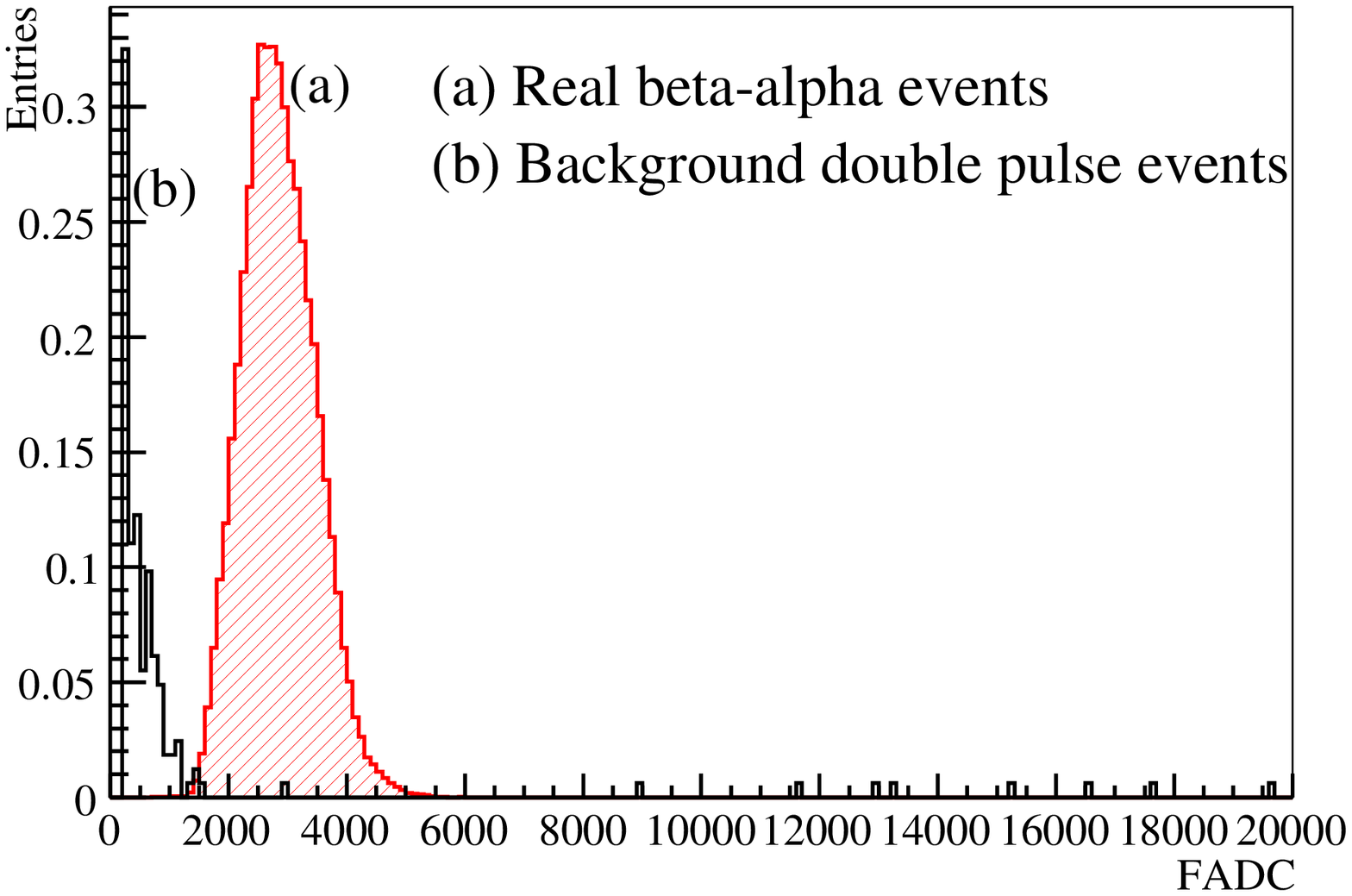}
\figcaption{\label{fig6}   The integral value of the second pulse (The entries are normalized)}
\end{center}

\subsubsection{Cuts selection}

According to the analysis of real $\beta$-$\alpha$ events, several cut criterions were selected.

(1)	The coincidence time window of the two PMTs was required to be smaller than 5 ns;

(2)	The difference of time interval of $\beta$-$\alpha$ cascade events detected by the two PMTs was within (-2 ns, 3 ns);

(3)	The integral value of $\alpha$ events was within (1700 FADC, 8000 FADC).

After bubbling $^{220}$Rn into LS for 20 hours, there were about 2$\times$10$^4$ $\beta$-$\alpha$ events detected in 17.10 g LS in 30 minutes' data taking after using the cut criterions discussed above.

\subsubsection{Background events}

Background events were taken for one day using the pure LS (17.10 g) without loading $^{220}$Rn to it. 163 double pulse events were found. But the integral value distribution of the second pulse (maybe the fake alpha event pulse) and the difference of time interval of the double pulse events detected by the two PMTs¡¯ were much different with the real $\beta$-$\alpha$ events, as shown in Fig.~\ref{fig5}(b) and Fig.~\ref{fig6}(b). The integral values of the second pulses were almost less than 1700 FADC and only a small percent of the difference of the time interval was within (-2 ns, 3 ns). After applying these cut criterions to background events, there remained only 2 background $\beta$-$\alpha$ cascade events a day, while there was 163 without cut selections.

Compared with 2$\times$10$^4$ real $\beta$-$\alpha$ cascade events detected in $^{220}$Rn loaded LS in 30 minutes, the 2 background events in one day can be neglected in the following study.

\subsection{The study of water extraction}

Purification by water extraction relies on the polarity of water molecules to separate polarized impurities, e.g. free-state ions of radioactive metals, from the non-polar LAB and fluor molecules. Water extraction is very effective for most ionic metals such as K, Ra, and Bi, but with some effectiveness for Po and Pb. For Po and Pb, the reduction was seen to be equally fast but less effective with an 82$\sim$87\% removal fraction in SNO+ laboratory study \cite{lab7} \cite{lab8}.

After the scintillator had a good $^{212}$Pb concentration, the scintillator was purified by water extraction. The purification efficiency was defined as,

\begin{eqnarray}
\label{eq2}
u = \frac{y-x}{y} = 1 - \frac{x}{y}.
\end{eqnarray}

Here, $x$ means event number after purification, $y$ means event number before purification and $u$ means purification efficiency.

The statistical error of efficiency is calculated by Clopper-Pearson parameter estimation of Binomial Distribution~\cite{zhuyongsheng}. The formula is described as below,

\begin{eqnarray}
\label{eq4}
\sigma_+=\big(1+\frac{n-\hat s}{\hat s+1}f_{1-\alpha/2}(2(n-\hat s),2(\hat s+1))\big)^{-1}-\hat p,\\
\sigma_-=\hat p-\big(1+\frac{n-\hat s+1}{\hat s}f_{\alpha/2}(2(n-\hat s+1),2\hat s)\big)^{-1}.
\end{eqnarray}

Here, 1-$\alpha$ is the confidence level; n is the total events number; $f_{\alpha/2}$ is the upside $\alpha/2$ fractile of F-distribution; $\hat s$ is the passed events number and $\hat p=\hat s/n$. In this paper, $n=y,~\hat s=y-x$ and the confidence level is set as 0.683.

Water extraction was done with equal amount of 12 M¦¸ deionized water and liquid scintillator. A separatory funnel and a magneton were used to mix 35 ml of water and 35 ml of scintillator. The solution was mixed and then separated. The operation was called one stage extraction. After each separation a clean separatory funnel and fresh deionized water were used to do multiple stages extraction. Scintillator samples were placed in small test tube which contained 17.1 g of LS sample. Then the $\beta$-$\alpha$ counting system was used to measure the radioactivity in scintillator before and after purification.

Stability study of the $\beta$-$\alpha$ counting system was conducted by measuring purification efficiency at three different time with the same LS sample. The efficiencies were consistent to each other, which were 84.3$^{+1.2}_{-1.3}$\%, 82.7$^{+1.5}_{-1.6}$\% and 83.3$^{+1.8}_{-1.9}$\%. Therefore, it was reliable to optimize purification parameters by using the $\beta$-$\alpha$ counting system.

Fig.~\ref{fig7} shows the relationship between extraction efficiency and the stirring time. The stirring speed was 600 r/min, while the stirring time was 1min, 2 min, 4 min, 8 min, 16 min and 32 min. The efficiency increased slowly after extraction for 8 min. When extracted for 32 min, the radioactivity of LS decreased 86.7$^{+0.5}_{-0.5}$\% with 737 $\beta$-$\alpha$ cascade events in LS samples. Before purification there were 5529 $\beta$-$\alpha$ cascade events. The error in Fig.7 to Fig.9 is statistical error.

Fig.~\ref{fig8} shows the relationship between extraction efficiency and the extraction stage. The extraction time was 3 min and the stirring speed was set at 1200 r/min. After extraction for 5 stages, the purification efficiency became almost stable, reaching a not very high efficiency of 92.1$^{+0.3}_{-0.4}$\%. Since Pb was a very polar atom, it was expected that its appetency to water will be much higher than LS of several orders of magnitude. The most likely explanation was that a fraction of the Pb was bound in a nonpolar configuration which reduced the partitioning coefficient and thus the purification efficiency \cite{lab7}.

Fig.~\ref{fig9} shows the relationship between extraction efficiency and the volume proportion of LS to water. In laboratory measurement, the purification efficiency decreased slowly when the proportion of LS to water was larger than 6. Then the efficiency decreased shapely as soon as the volume proportion reached 6. According to the result, the volume proportion of LS to water used in JUNO purification can be set at 5.

\begin{center}
\includegraphics[width=7cm]{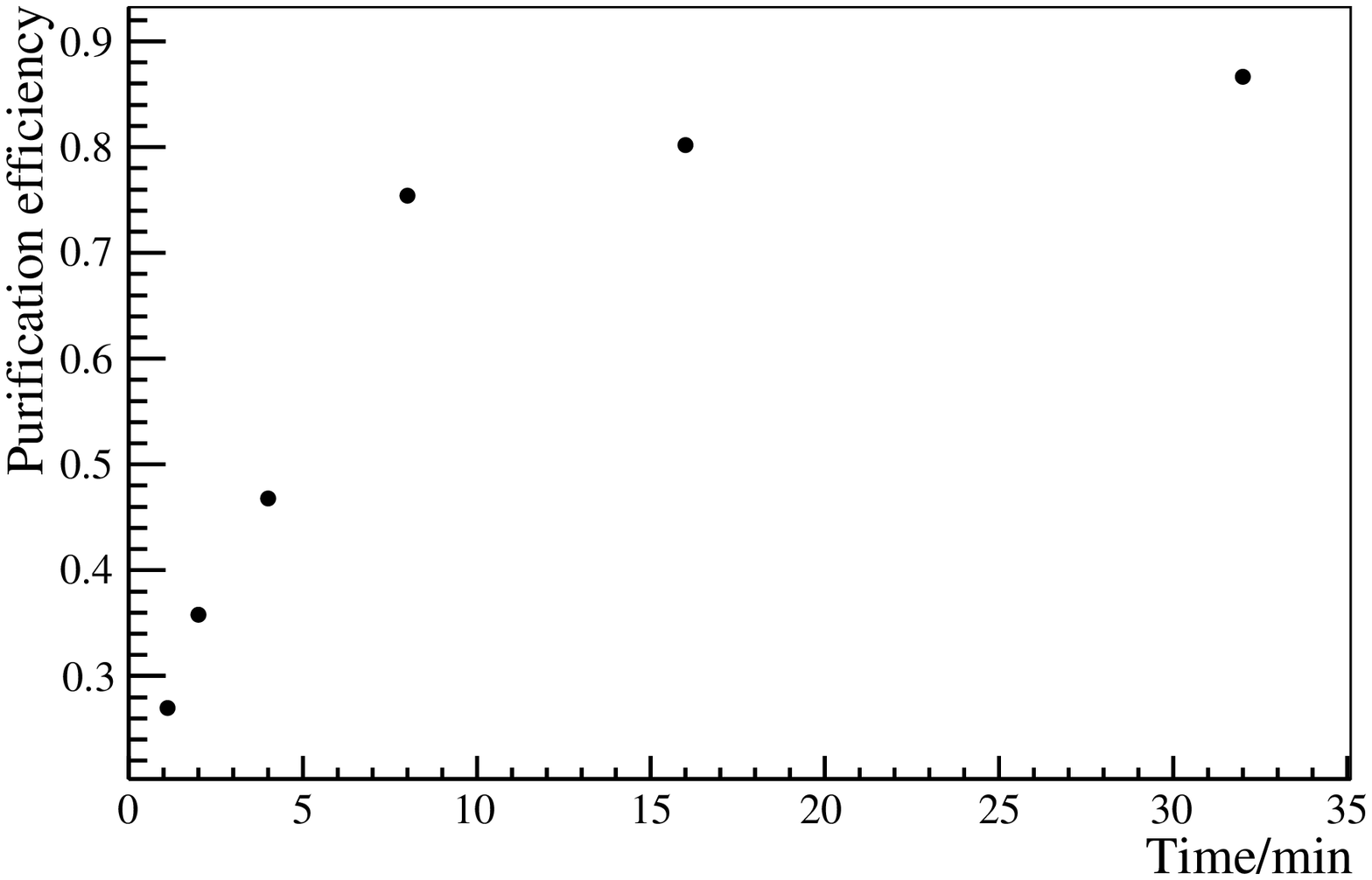}
\figcaption{\label{fig7}   Purification efficiency VS extraction time, 600 r/min stirring speed }
\end{center}

\begin{center}
\includegraphics[width=7cm]{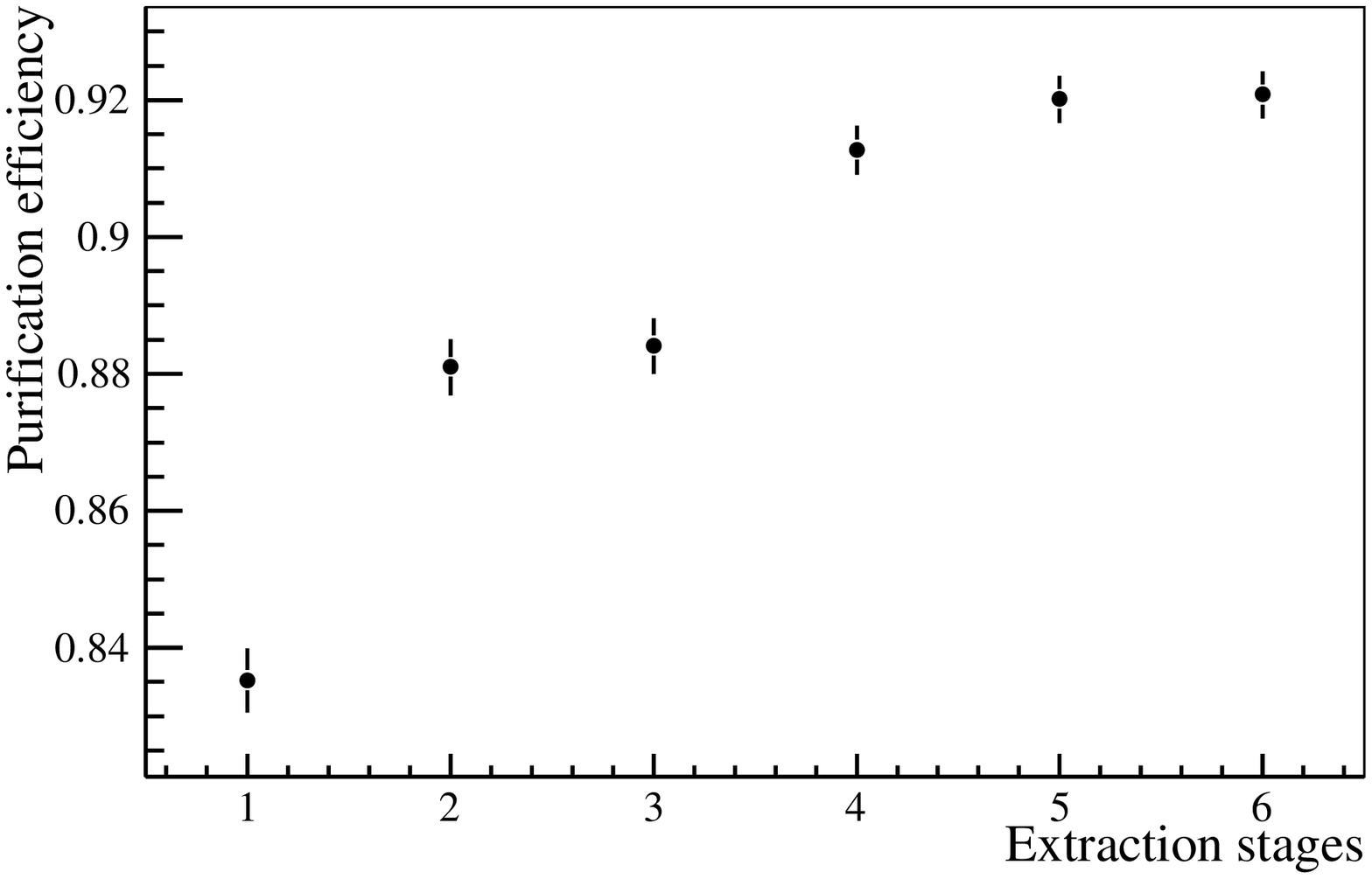}
\figcaption{\label{fig8}   Purification efficiency VS extraction stages, 3 min extraction time and 1200 r/min stirring speed }
\end{center}

\begin{center}
\includegraphics[width=7cm]{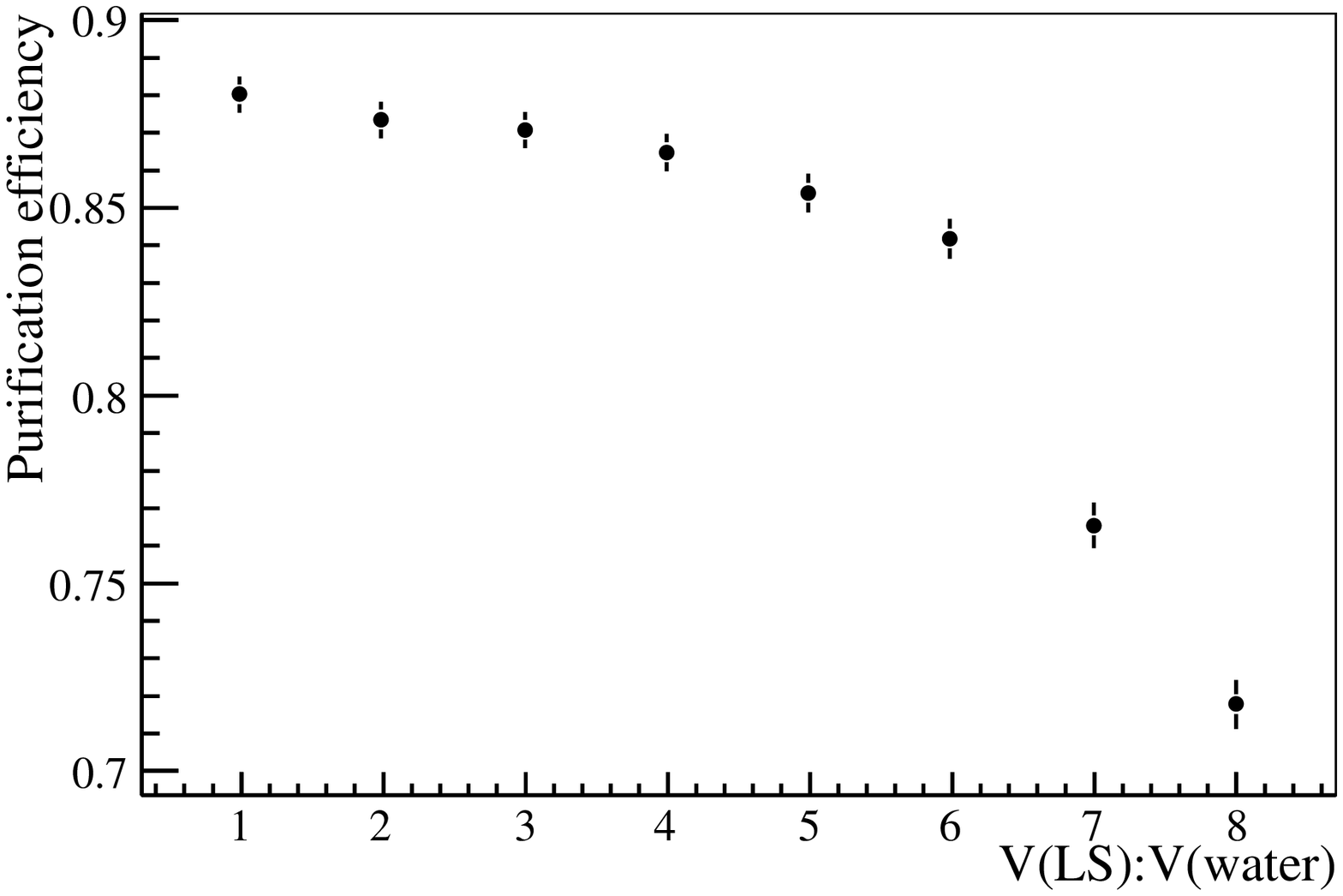}
\figcaption{\label{fig9}   Purification efficiency VS the volume proportion of LS and water, 10 min extraction time and 800 r/min stirring speed}
\end{center}

The water extraction efficiency for $^{212}$Pb can reach more than 84\% with laboratory scale purification setup. The extraction stage and volume ratio of LS to water can be set at 5 in future large scale purification plants design and operation.

\section{Conclusions}

To study the water extraction in the future JUNO LS purification plants, an extraction prototype had been constructed, and a background free efficiency measurement had been achieved with $^{220}$Rn loaded LS. The measured water extraction efficiency to $^{212}$Pb was about no less than 84\%, reaching the world average level, and optimized operation parameters had been obtained. Now, a medium scale water extraction tower had been built which was based on laboratory study results. The radioactivity loading setup and $\beta$-$\alpha$ counting system will be useful to the investigation of the parameters involved in large scale purification plants in the future.

\end{multicols}

\begin{multicols}{2}
\vspace{-1mm}
\centerline{\rule{80mm}{0.1pt}}

\end{multicols}

\clearpage

\end{document}